\documentclass[10pt,conference]{IEEEtran}

\usepackage{amsfonts}
\usepackage{amssymb}
\usepackage{mathrsfs}
\usepackage{verbatim}
\usepackage{times}
\usepackage{subfigure}
\usepackage[centertags]{amsmath}
\usepackage{graphicx}
\newtheorem{defn}{Definition}
\newtheorem{eg}{Example}

\begin{document}

\title{Algebraic Distributed Space-Time Codes with Low ML Decoding Complexity}

\author{
\authorblockN{G. Susinder Rajan}
\authorblockA{ECE Department \\
Indian Institute of Science\\
Bangalore 560012, India\\
susinder@ece.iisc.ernet.in}
\and
\authorblockN{B. Sundar Rajan}
\authorblockA{ECE Department\\
Indian Institute of Science\\
Bangalore 560012, India\\
bsrajan@ece.iisc.ernet.in}
}
%
\maketitle
\begin{abstract}
"Extended Clifford algebras" are introduced as a means to obtain low ML decoding complexity space-time block codes. Using left regular matrix representations of two specific classes of extended Clifford algebras, two systematic algebraic constructions of full diversity Distributed Space-Time Codes (DSTCs) are provided for any power of two number of relays. The left regular matrix representation has been shown to naturally result in space-time codes meeting the additional constraints required for DSTCs. The DSTCs so constructed have the salient feature of reduced Maximum Likelihood (ML) decoding complexity. In particular, the ML decoding of these codes can be performed by applying the lattice decoder algorithm on a lattice of four times lesser dimension than what is required in general. Moreover these codes have a uniform distribution of power among the relays and in time, thus leading to a low Peak to Average Power Ratio at the relays.
\end{abstract}

\section{Introduction}
\label{sec1}
  
Coding for wireless relay networks has received a lot of attraction recently with the advent of cooperative diversity techniques. In this paper, we are interested in constructing Distributed Space-Time Codes (DSTCs) for the Amplify and Forward (AF) based cooperative diversity protocol proposed by Jing and Hassibi \cite{JiH1}. The Jing and Hassibi protocol is a two phase based AF protocol. In the first phase, the source broadcasts a vector to all the $R$ relays which contains the information that the source intends to communicate to the destination. In the second phase, each relay transmits a vector obtained by linear processing of the received vector and its conjugate to the destination. To the destination, this would appear as if each relay transmitted a column of a linear space-time code thus leading to the concept of DSTCs. We refer the readers to \cite{JiH1} for a detailed introduction to DSTCs.
Consider a $R\times R$ linear design $S(X)$ in $2K$ real variables $x_1,x_2,\dots,x_{2K}$ as follows
$$
S(X)=\sum_{i=1}^{2K}x_iC_i
$$
where, the complex matrix $C_i\in {\mathbb C}^{R\times R}$ is called the \textit{'weight matrix'} corresponding to the real variable $x_i$ and $R$ is the number of relays. Let $X=\left[\begin{array}{cccc}x_1 & x_2 & \dots & x_{2K}\end{array}\right]^T$. From the $2K$ real variables we can form $K$ complex variables $z_1,z_2,\cdots,z_K$ by pairing two real variables at a time. Let $s=\left[\begin{array}{cccc}z_1 & z_2 & \dots & z_{K}\end{array}\right]^T.$ Then the linear design $S$ can also be expressed as 
\begin{equation}
S(X)=\left[\begin{array}{cccc}A_1s+B_1s^* & A_2s+B_2s^* & \dots & A_Rs+B_Rs^*\end{array}\right]
\end{equation}
where, $A_i,B_i \in {\mathbb C}^{R\times R},~~ i=1,\dots,R$ are complex matrices. We call the matrices $A_i,B_i$ as \textit{'relay matrices'}. In \cite{JiH1}, it has been shown that linear designs satisfying the following conditions
\begin{enumerate}
\item For any $i=1,\dots,R$, either $A_i=\bf{0}$ or $B_i=\bf{0}$ (Conjugate-Linearity Property) and $K=R$
\item All the nonzero relay matrices are unitary matrices
\end{enumerate}
\noindent
are applicable as DSTCs where $s=\left[\begin{array}{cccc}z_1 & z_2 & \dots & z_R\end{array}\right]^T$ will be the vector transmitted by the source in the first phase and the matrices $A_i,B_i$ will be used at the $i$th relay to perform linear processing of the received vector and its conjugate.

All the previous works on DSTC construction except \cite{KiR1}, \cite{JiJ}, \cite{RaR2} and \cite{RaR3} do not address the important problem of designing DSTCs with low Maximum Likelihood (ML) decoding complexity. This problem gains significant importance especially if the number of relays in the network is large. Suppose we partition the $2K$ real variables and their corresponding weight matrices into $g$-groups $L_k,k=1,\dots,g$, the $k$-th group containing $2K/g$ real variables. Then without loss of generality we can consider the natural simplest partition, in terms of which $S(X)$ can be written as,
$$
S(X)=\sum_{k=1}^{g}S_k(X_k)\quad \mathrm{where},\ S_k(X_k)=\sum_{i=\frac{(k-1)2K}{g}+1}^{\frac{2kK}{g}}x_iC_i.
$$
If the received matrix and the channel matrix are denoted by $Y$ and $H$ respectively, then a ML decoder in general minimizes the metric $\parallel Y-S(X)H\parallel^2$. However, if 
\begin{equation}
\label{eqn_4gp}
C_i^HC_j+C_j^HC_i=\mathbf{0},\forall i\in L_p,\forall j\in L_q,p\neq q, 1\leq p,q\leq g
\end{equation}
then this is equivalent \cite{RaR2} to minimizing
\begin{equation}
\label{eqn_submetric}
\parallel Y-S_k(X_k)H\parallel^2
\end{equation}
for each $1\leq k\leq g$ individually. Note that \eqref{eqn_submetric} can be computed by applying the lattice decoder algorithm on a lattice of $g$ times lesser dimension. We then say that the DSTC is $g$-group ML decodable. In \cite{KiR1}, the authors constructed $2$-group ML decodable codes using division algebras. In \cite{JiJ}, DSTCs for two and four relays based on the Alamouti design and quasi-orthogonal design were proposed. Following the works of \cite{KiR1}, in \cite{RaR2}, a class of $4$-group ML decodable codes called Precoded Co-ordinate Interleaved Orthogonal Designs (PCIODs) were constructed for arbitrary number of relays. However, PCIODs have a drawback that the power distribution among the relays is not uniform across time slots thus leading to a large Peak to Average Power Ratio (PAPR). Moreover the relay matrices of PCIODs are not unitary which in turn forces the destination to perform additional processing \cite{RaR2} to make the covariance matrix of the resultant noise vector at the destination a scaled identity matrix. Recently in \cite{RaR3}, this problem was resolved by an alternative iterative construction for number of relays $R$ being a power of two. However, both the constructions of \cite{RaR2,RaR3} were not obtained from a systematic algebraic procedure targeting the requirements for low decoding complexity.
The main contributions of this paper are as follows.
\begin{itemize}
\item A generalization of Clifford algebras, which we call  "Extended Clifford Algebras" is introduced as an algebraic framework to handle the problem of constructing STBCs with low ML decoding complexity. To the knowledge of the authors, this is the first known systematic algebraic procedure to solve this problem. This algebraic framework simplifies the problem to finding appropriate matrix representations of extended Clifford algebras.
\item Using left regular representation of "Extended Clifford Algebras", two different fully diverse algebraic DSTC constructions are provided for power of two number of relays. Left regular representation has been shown to naturally result in space-time codes meeting the additional requirements of DSTCs. Moreover, one of the constructions provides an algebraic explanation for the recently proposed DSTC construction in \cite{RaR3}. 
\item ML decoding of these algebraic DSTCs can be performed by applying the well known lattice decoder algorithm on a lattice of four times lesser dimension than what is required in general.
\item Furthermore, the proposed DSTCs have lesser PAPR than the DSTCs of similar ML decoding complexity reported in \cite{RaR2}. 
\end{itemize}

The rest of the paper is organized as follows: In Section \ref{sec2} we briefly recollect a set of known sufficient conditions for low ML decoding complexity designs. Extended Clifford algebras are introduced in Section \ref{sec3} and algebraic construction of $4$-group ML decodable DSTCs is presented in a general setting. The two special classes of codes from extended Clifford algebras are presented in detail in Sections \ref{sec4} and \ref{sec5}. In Section \ref{sec6} it is shown that all the requirements for DSTCs are satisfied by the proposed codes.

\noindent
\textbf{Notation:}
For a complex matrix $A$, $A_{I}$ denotes the real matrix obtained by taking the real parts of all the entries of $A$ and $A_{Q}$ denotes the real matrix obtained by taking the imaginary parts of all the entries of $A$. If ${\cal A}$ is an algebra over a field $F$ then $End_F({\cal A})$ denotes the set of all maps from ${\cal A}$  to ${\cal A}$ that are $F-$linear.

\section{Sufficient conditions for low ML decoding complexity designs}
\label{sec2}
Recently in \cite{KaR3}, sufficient conditions for designing $g$-group ML decodable STBCs have been reported. Since our constructions rely upon these sufficient conditions and because we restrict ourselves to the $g=4$ case, we briefly introduce these sufficient conditions \cite{KaR3} for $g=4$ before proceeding further. Essentially we would like to be able to partition the set of weight matrices of a linear design into $4$ groups in such a way that the condition in \eqref{eqn_4gp} is satisfied. Let us first list down the $K$ weight matrices in the form of an array as shown below.
\begin{center}
\begin{tabular}{c|ccc}
$C_1$ & $C_{\frac{K}{4}+1}$ & $C_{\frac{K}{2}+1}$ & $C_{\frac{3K}{4}+1}$\\
\hline
$C_2$ & $C_{\frac{K}{4}+2}$ & $C_{\frac{K}{2}+2}$ & \\
$\vdots$ & $\vdots$ & $\ddots$ & $\vdots$\\
$C_{\frac{K}{4}}$ & $C_{\frac{K}{2}}$ & $C_{\frac{3K}{4}}$ & $C_{K}$\\
\end{tabular}
\end{center}
The partitioning is as follows: All the weight matrices in one column belong to one group. To simplify the construction, we shall consider all the weight matrices to be unitary and furthermore set $C_1=I$. Then it has been shown in \cite{KaR3} that it is sufficient to design the matrices in the first row and the first column such that they satisfy the following conditions. 
\begin{enumerate}
\item All the matrices in the first row except $C_1=I$ should square to $-I$ and should pair-wise anti-commute among themselves. 
\item The matrices in the first column should square to $I$ and should commute with all the matrices in the first row and the first column.
\end{enumerate}
Once such a set of matrices is obtained, the matrix in the $i$-th row and $j$-th column can be filled up by multiplying $C_i$ and $C_{\frac{(j-1)K}{4}+1}$. It can be easily verified that such a set of weight matrices will satisfy the conditions in \eqref{eqn_4gp} for $g=4$.

\section{Algebraic construction $4$-group ML decodable DSTCs}
\label{sec3}
 An algebra is simply a ring as well as a vector space with the addition operation being compatible to both the ring and the vector space structures. In this section, we introduce the algebraic framework of "Extended Clifford Algebras" to handle the problem of constructing $g$-group ML decodable codes satisfying the sufficient conditions discussed in the previous section \cite{RaR3}. Using left regular representation of extended Clifford algebras we then obtain two constructions of $4$-group ML decodable DSTCs. 

Our methodology to construct the matrices in the first row and first column (as discussed in previous section) would be to fabricate an algebra in such a way that it contains elements satisfying the algebraic relations we need. Once we construct the algebra, we then obtain the required linear design by taking an appropriate matrix representation of the constructed algebra.
\begin{defn}\cite{TiH}
The Clifford algebra, denoted by $Cliff_n$ is the algebra over the real field $\mathbb{R}$ generated by $n$ objects $\gamma_k,\ k=1,\dots,n$ which are anti-commuting ($\gamma_k\gamma_j=-\gamma_j\gamma_k,\ \forall k\neq j$) and squaring to $-1$ ($\gamma_k^2=-1\ \forall k=1,\dots,n$).
\end{defn}
A natural basis for $Cliff_n$ seen as a vector space over $\mathbb{R}$~ is 
\begin{equation}
\begin{array}{rl}
\mathscr{B}_n=&\left\{1\right\}\bigcup\left\{\gamma_i|i=1,\dots,n\right\}\\
&\bigcup_{m=2}^{n}\left\{\prod_{i=1}^{m}\gamma_{k_i}|1\leq k_i\leq k_{i+1}\leq n\right\}
\end{array}
\end{equation}
The number of basis elements is $|\mathscr{B}_n|=2^n$.

Notice that the defining algebraic relations of the generators of a Clifford algebra resemble the algebraic relations which the matrices in the first row should satisfy. Hence we can obtain the matrices in the first row by taking unitary matrix representations of the generators of a Clifford algebra. To obtain the matrices in the first column, we use a similar strategy. We introduce few new symbols in the Clifford algebra and define them to square to $1$ and commute with the generators of the Clifford algebra and also commute among themselves. In other words, after introducing new symbols, multiplication in the algebra is appropriately defined in order to create a bigger algebra which contains the Clifford algebra as a sub-algebra. Hence by taking the unitary matrix representation of these specific elements of the algebra, we get the weight matrices of the required linear STBC. 
\begin{defn}
Let $L=2^a,a\in \mathbb{N}$. An Extended Clifford algebra denoted by $\mathbb{A}_n^L$ is the associative algebra over $\mathbb{R}$~ generated by $n+a$ objects $\gamma_k,\ k=1,\dots,n$ and $\delta_i,\ i=1,\dots,a$ which satisfy the following relations:
\begin{itemize}
\item $\gamma_k^2=-1,\ \forall\ k=1,\dots,n$
\item $\gamma_k\gamma_j=-\gamma_j\gamma_k,\ \forall\ k\neq j$
\item $\delta_k^2=1,\ \forall k=1,\dots,a$
\item $\delta_k\delta_j=\delta_j\delta_k,\ \forall\ 1\leq k,j\leq a$
\item $\delta_k\gamma_j=\gamma_j\delta_k,\ \forall\ 1\leq k\leq a, 1\leq j\leq n$
\end{itemize}
\end{defn}

From the above definition, it is clear that $Cliff_n$ is a sub-algebra of $\mathbb{A}_n^L$. Let $\mathscr{B}_n$ be the natural $\mathbb{R}$~ basis for this sub-algebra $Cliff_n$. Then a natural $\mathbb{R}$~ basis for $\mathbb{A}_n^L$ is given by 
\begin{equation}
\begin{array}{rl}
\mathscr{B}_n^L=&\mathscr{B}_n\cup\left\{\mathscr{B}_n\delta_i|i=1,\dots,a\right\}\\
&\bigcup_{m=2}^{a}\mathscr{B}_n\left\{\prod_{i=1}^{m}\delta_{k_i}|1\leq k_i\leq k_{i+1}\leq a\right\}.
\end{array}
\end{equation}
Thus the dimension of $\mathbb{A}_n^L$ seen as a vector space over $\mathbb{R}$ is $2^{n+a}$.

\begin{eg}
Let us take $n=2$, $a=1$. Hence $L=2$. Then 
$$
\mathbb{A}_2^2=\left\{a_1+\gamma_1a_2+\delta_1a_3+\delta_1\gamma_1a_4|a_1,a_2,a_3,a_4\in\mathbb{R}\right\}.
$$
Addition in the algebra is defined to be component wise and multiplication is completely described by defining the multiplication between any two basis elements. The multiplication table can be easily generated using the defining algebraic relations of the generators and is given as follows.
\begin{center}
\begin{tabular}{|c|c|c|c|c|}
\hline
~ & $1$ & $\gamma_1$ & $\delta_1$ & $\delta_1\gamma_1$\\
\hline
$1$ & $1$ & $\gamma_1$ & $\delta_1$ & $\delta_1\gamma_1$\\
\hline
$\gamma_1$ & $\gamma_1$ & $-1$ & $\delta_1\gamma_1$ & $-\delta_1$\\
\hline
$\delta_1$ & $\delta_1$ & $\delta_1\gamma_1$ & $1$ & $\gamma_1$\\
\hline
$\delta_1\gamma_1$ & $\delta_1\gamma_1$ & $-\delta_1$ & $\gamma_1$ & $-1$\\
\hline
\end{tabular}
\end{center} 
One can check from the multiplication table that the multiplication is indeed associative. Note that $A_2^2$ can also be viewed as a vector space over $\mathbb{C}$ by thinking of the symbol $\gamma_1$ as the complex number $i=\sqrt{-1}$. Then, we have
$$
\mathbb{A}_2^2=\left\{z_1+\delta_1z_2|z_1,z_2\in\mathbb{C}\right\}
$$
where, $z_1=a_1+\gamma_1a_2$ and $z_2=a_3+\gamma_1a_4$
\end{eg}

Since we are interested in $4$-group decodable DSTCs, we need $4$ matrices (including identity matrix) in the first row. One way to obtain such matrices is to take the matrix representation of $\mathbb{A}_3^L$ for $L=2^a,a\in\mathbb{N}$. The matrix representation of the symbols $1,\gamma_1,\gamma_2,\gamma_3$ respectively can be used to fill up the first row. Interestingly, there is yet another way of obtaining such matrices. Let us look at $\mathbb{A}_2^L$ for $L=2^a,a\in \mathbb{N}$. The symbols $\gamma_1$ and $\gamma_2$ square to $-1$ and anticommute. However note that 

{\small
\begin{equation*}
(\gamma_2\gamma_1)^2=-1;~  (\gamma_2\gamma_1)\gamma_1=-\gamma_1(\gamma_2\gamma_1);~ 
(\gamma_2\gamma_1)\gamma_2=-\gamma_2(\gamma_2\gamma_1).
\end{equation*}
}

Thus the symbol $\gamma_2\gamma_1$ also squares to $-1$ and anticommutes with the symbols $\gamma_1$ and $\gamma_2$. Thus we can fill up the first row with the matrix representations of the symbols $1,\gamma_1,\gamma_2,\gamma_2\gamma_1$ respectively. Thus we get two classes of $4$-group ML decodable STBCs, one from $\mathbb{A}_3^L$ and the other from $\mathbb{A}_2^L$.

\subsection{Matrix Representation}
There are several ways to obtain a matrix representation of an algebra. However we need to take an appropriate matrix representation such that the following conditions are satisfied.
\begin{enumerate}
\item The symbols $1$, $\gamma_1$, $\gamma_2$, $\dots$, $\gamma_n$, $\delta_k,k=1,\dots,a$, $\bigcup_{m=2}^{a}\prod_{i=1}^{m}\delta_{k_i}|1\leq k_i\leq k_{i+1}\leq a$ should be represented by unitary matrices.
\item The resulting linear design should have the Conjugate-Linearity property.  
\item All the relay matrices should be unitary.
\end{enumerate}

Such matrices are naturally provided by the left regular representation of the associative algebra $\mathbb{A}_n^L$. Left regular representation is an easy way to obtain the matrix representation for any finite dimensional associative algebra \cite{Jac}. The first requirement of unitary matrix representation is met because the natural basis elements of $\mathbb{A}_n^L$ together with their negatives form a finite group under multiplication. This fact in conjunction with the properties of left regular representation guarantee a unitary matrix representation for the required symbols. We shall prove the other properties in Section \ref{sec6} after illustrating the construction procedure for both the codes from $\mathbb{A}_2^L$ as well as those from $\mathbb{A}_3^L$.

\section{Codes from $\mathbb{A}_2^L$}
\label{sec4}
We first view $\mathbb{A}_2^L$ as a vector space over $\mathbb{C}$ by thinking of $\gamma_1$ as the complex number $i=\sqrt{-1}$. A natural $\mathbb{C}$~ basis for $\mathbb{A}_2^L$ is given by

\begin{equation*}
\begin{array}{rl}
\mathcal{B}_n^L=&\left\{1,\gamma_2\right\}\cup\left\{\left\{1,\gamma_2\right\}\delta_i|i=1,\dots,a\right\}\\
& \bigcup_{m=2}^{a}\left\{1,\gamma_2\right\}\left\{\prod_{i=1}^{m}\delta_{k_i}|1\leq k_i\leq k_{i+1}\leq a\right\}
\end{array}
\end{equation*}
The dimension of $\mathbb{A}_2^L$ seen as a vector space over $\mathbb{C}$ is $2^{n+a-1}$.

We have a natural embedding of $\mathbb{A}_2^L$ into $\mathrm{End}_{\mathbb{C}}(\mathbb{A}_2^L)$ given by left multiplication \cite{Jac} as shown below:
\begin{equation*}
\phi:\mathbb{A}_2^L\mapsto \mathrm{End}_{\mathbb{C}}(\mathbb{A}_2^L);~~~
\phi(x)=L_x: y\mapsto xy.
\end{equation*}

Since the map $L_x$ is $\mathbb{C}$~-linear, we can write down a matrix representation of $L_x$ with respect to the natural $\mathbb{C}$~ basis $\mathcal{B}_n^L$. Thus we obtain a design satisfying the requirements of \eqref{eqn_4gp} for $g=4$.
\begin{eg}
Let us begin with $N_T=2$ transmit antennas. Let $n=2$. Then equating $n+a-1=1$, we get $a=0$ and hence $L=1$. But the algebra $\mathbb{A}_2^1$ is same as $Cliff(2)$ which is nothing but the Hamiltonian Quaternions $\mathbb{H}$. It is well known \cite{Jac} that the left regular matrix representation of $\mathbb{H}$ yields the popular Alamouti design. Thus we see that our algebraic code construction which was driven by the need for low ML decoding complexity naturally leads to the Alamouti design.
\end{eg}
\begin{eg}
Suppose we want a design for $N_T=8=2^3$ transmit antennas. Let $n=2$. Then we need $n+a-1=3$. Thus $a=2$ and $L=4$. A general element of the algebra $\mathbb{A}_2^4$ looks like

{\small
$$
x=z_1+\delta_1z_2+\delta_2z_3+\delta_1\delta_2z_4+\gamma_2z_5+\delta_1\gamma_2z_6+\delta_2\gamma_2z_7+\delta_1\delta_2\gamma_2z_8
$$
}

\noindent
where, $z_i\in\mathbb{C},\forall i=1,\dots,8$. The image of the basis $\mathcal{B}_2^4$ under the map $\phi$ is shown in \eqref{eqn_matrix} at the top of the next page. Thus, we have
\begin{figure*}  
\begin{equation}
\label{eqn_matrix}
\begin{array}{rcl}
\phi(1)&=&z_1+\delta_1z_2+\delta_2z_3+\delta_1\delta_2z_4+\gamma_2z_5+\delta_1\gamma_2z_6+\delta_2\gamma_2z_7+\delta_1\delta_2\gamma_2z_8\\
\phi(\delta_1)&=&\delta_1z_1+z_2+\delta_1\delta_2z_3+\delta_2z_4+\delta_1\gamma_2z_5+\gamma_2z_6+\delta_1\delta_2\gamma_2z_7+\delta_2\gamma_2z_8\\
\phi(\delta_2)&=&\delta_2z_1+\delta_1\delta_2z_2+z_3+\delta_1z_4+\delta_2\gamma_2z_5+\delta_1\delta_2\gamma_2z_6+\gamma_2z_7+\delta_1\gamma_2z_8\\
\phi(\delta_1\delta_2)&=&\delta_2\delta_2z_1+\delta_2z_2+\delta_1z_3+z_4+\delta_1\delta_2\gamma_2z_5+\delta_2\gamma_2z_6+\delta_1\gamma_2z_7+\gamma_2z_8\\
\phi(\gamma_2)&=&\left(z_1+\delta_1z_2+\delta_2z_3+\delta_1\delta_2z_4+\gamma_2z_5+\delta_1\gamma_2z_6+\delta_2\gamma_2z_7+\delta_1\delta_2\gamma_2z_8\right)\gamma_2\\
&=&\gamma_2z_1^*+\delta_1\gamma_2z_2^*+\delta_2\gamma_2z_3^*+\delta_1\delta_2\gamma_2z_4^*-z_5^*-\delta_1z_6^*-\delta_2z_7^*-\delta_1\delta_2z_8^*\\
\phi(\delta_1\gamma_2)&=&\delta_1\gamma_2z_1^*+\gamma_2z_2^*+\delta_1\delta_2\gamma_2z_3^*+\delta_2\gamma_2z_4^*-\delta_1z_5^*-z_6^*-\delta_1\delta_2z_7^*-\delta_2z_8^*\\
\phi(\delta_2\gamma_2)&=&\delta_2\gamma_2z_1^*+\delta_1\delta_2\gamma_2z_2^*+\gamma_2z_3^*+\delta_1\gamma_2z_4^*-\delta_2z_5^*-\delta_2z_6^*-z_7^*-\delta_1z_8^*\\
\phi(\delta_1\delta_2\gamma_2)&=&\delta_1\delta_2\gamma_2z_1^*+\delta_2\gamma_2z_2^*+\delta_1\gamma_2z_3^*+\gamma_2z_4^*-\delta_1\delta_2z_5^*-\delta_2z_6^*-\delta_1z_7^*-z_8^*
\end{array}
\end{equation}
\hrule
\end{figure*}

{\small
\begin{equation*}
\label{eqn_design8}
L_x=\left[\begin{array}{ccccrrrr}
z_1 & z_2 & z_3 & z_4 & -z_5^* & -z_6^* & -z_7^* & -z_8^*\\
z_2 & z_1 & z_4 & z_3 & -z_6^* & -z_5^* & -z_8^* & -z_7^*\\
z_3 & z_4 & z_1 & z_2 & -z_7^* & -z_8^* & -z_5^* & -z_6^*\\
z_4 & z_3 & z_2 & z_1 & -z_8^* & -z_7^* & -z_6^* & -z_5^*\\
z_5 & z_6 & z_7 & z_8 & z_1^* & z_2^* & z_3^* & z_4^*\\
z_6 & z_5 & z_8 & z_7 & z_2^* & z_1^* & z_4^* & z_3^*\\
z_7 & z_8 & z_5 & z_6 & z_3^* & z_4^* & z_1^* & z_2^*\\
z_8 & z_7 & z_6 & z_5 & z_4^* & z_3^* & z_2^* & z_1^*
\end{array}\right].
\end{equation*}
}
Also, we have

{\small
\begin{equation*}
\begin{array}{rcl}
x&=&z_{1I}+\gamma_1z_{1Q}+\delta_1z_{2I}+\delta_1\gamma_1z_{2Q}\\
&&+\delta_2z_{3I}+\delta_2\gamma_1z_{3Q}+\delta_1\delta_2z_{4I}+\delta_1\delta_2\gamma_1z_{4Q}\\
&&+\gamma_2z_{5I}+\gamma_2\gamma_1z_{5Q}+\delta_1\gamma_2z_{6I}+\delta_1\gamma_2\gamma_1z_{6Q}\\
&&+\delta_2\gamma_2z_{7I}+\delta_2\gamma_2\gamma_1z_{7Q}+\delta_1\delta_2\gamma_2z_{8I}+\delta_1\delta_2\gamma_2\gamma_1z_{8Q}
\end{array}
\end{equation*}
leading to
\begin{equation*}
\label{eqn_wtmat}
\begin{array}{rl}
L_x=&\phi(1)z_{1I}\phi(1)+\phi(\gamma_1)z_{1Q}+\phi(\delta_1)z_{2I}\\
&+\phi(\delta_1\gamma_1)z_{2Q}+\phi(\delta_2)z_{3I}+\phi(\delta_2\gamma_1)z_{3Q}\\
&+\phi(\delta_1\delta_2)z_{4I}+\phi(\delta_1\delta_2\gamma_1)z_{4Q}+\phi(\gamma_2)z_{5I}\\
&+\phi(\gamma_2\gamma_1)z_{5Q}+\phi(\delta_1\gamma_2)z_{6I}+\phi(\delta_1\gamma_2\gamma_1)z_{6Q}\\
&+\phi(\delta_2\gamma_2)z_{7I}+\phi(\delta_2\gamma_2\gamma_1)z_{7Q}+\phi(\delta_1\delta_2\gamma_2)z_{8I}\\
&+\phi(\delta_1\delta_2\gamma_2\gamma_1)z_{8Q}
\end{array}
\end{equation*}
}

\noindent
which explicitly gives the design $L_x$ in terms of its weight matrices. Expressing the elements of the algebra, the real variables of the resulting design and their corresponding weight matrices in the form of a tabular column as discussed in Section \ref{sec2}, we get

{\footnotesize
\begin{center}
\begin{tabular}{|c|c|c|c|}
\hline
$1$ & $\gamma_1$ & $\gamma_2$ & $\gamma_2\gamma_1$\\
$\phi(1)$ & $\phi(\gamma_1)$ & $\phi(\gamma_2)$ & $\phi(\gamma_2\gamma_1)$\\
$z_{1I}$ & $z_{1Q}$ & $z_{5I}$ & $z_{5Q}$\\
\hline
$\delta_1$ & $\delta_1\gamma_1$ & $\delta_1\gamma_2$ & $\delta_1\gamma_2\gamma_1$\\
$\phi(\delta_1)$ & $\phi(\delta_1)\phi(\gamma_1)$ & $\phi(\delta_1)\phi(\gamma_2)$ & $\phi(\delta_1)\phi(\gamma_2\gamma_1)$\\
$z_{2I}$ & $z_{2Q}$ & $z_{6I}$ & $z_{6Q}$\\
\hline
$\delta_2$ & $\delta_2\gamma_1$ & $\delta_2\gamma_2$ & $\delta_2\gamma_2\gamma_1$\\
$\phi(\delta_2)$ & $\phi(\delta_2)\phi(\gamma_1)$ & $\phi(\delta_2)\phi(\gamma_2)$ & $\phi(\delta_2)\phi(\gamma_2\gamma_1)$\\
$z_{3I}$ & $z_{3Q}$ & $z_{7I}$ & $z_{7Q}$\\
\hline
$\delta_1\delta_2$ & $\delta_1\delta_2\gamma_1$ & $\delta_1\delta_2\gamma_2$ & $\delta_1\delta_2\gamma_2\gamma_1$\\
$\phi(\delta_1\delta_2)$ & $\phi(\delta_1\delta_2)\phi(\gamma_1)$ & $\phi(\delta_1\delta_2)\phi(\gamma_2)$ & $\phi(\delta_1\delta_2)\phi(\gamma_2\gamma_1)$\\
$z_{4I}$ & $z_{4Q}$ & $z_{8I}$ & $z_{8Q}$\\
\hline
\end{tabular}
\end{center}
}
From the table above, it is clear how the weight matrices and real variables can be partitioned into four groups.
\end{eg}

In general for $N_T=2^\lambda$ transmit antennas we take the left regular representation of $\mathbb{A}_2^{2^{\lambda-1}}$ to obtain a $4$-group ML decodable linear design satisfying \eqref{eqn_4gp} for $g=4$. These codes were first obtained using a non-algebraic iterative construction procedure in \cite{RaR3}. The algebraic framework presented here provides an interesting algebraic explanation for the codes in \cite{RaR3}. 

\section{Codes from $\mathbb{A}_3^L$}
\label{sec5}
We use a slightly different approach to obtain codes from $\mathbb{A}_3^L$. Let us first consider the algebra, $\mathbb{A}_3^1$ which is nothing but $Cliff_3$. A general element of $Cliff_3$ looks like

{\small
\begin{equation*}
x=\hat{a}_1+\gamma_1\hat{a}_2+\gamma_2\hat{a}_3+\gamma_3\hat{a}_4+\gamma_1\gamma_2\hat{a}_5+\gamma_2\gamma_3\hat{a}_6+\gamma_1\gamma_3\hat{a}_7+\gamma_1\gamma_2\gamma_3\hat{a}_8
\end{equation*}
}

\noindent
for some $\hat{a}_i\in\mathbb{R},i=1,\dots,8$.  The element $\gamma_1\gamma_2\gamma_3$ satisfies the following properties.
{\small
\begin{equation*}
\begin{array}{rl}
(\gamma_1\gamma_2\gamma_3)^2=1; & (\gamma_1)(\gamma_1\gamma_2\gamma_3)=(\gamma_1\gamma_2\gamma_3)(\gamma_1);\\
(\gamma_2)(\gamma_1\gamma_2\gamma_3)=(\gamma_1\gamma_2\gamma_3)(\gamma_2); & 
(\gamma_3)(\gamma_1\gamma_2\gamma_3)=(\gamma_1\gamma_2\gamma_3)(\gamma_3).
\end{array}
\end{equation*}
}

Thus the element $\gamma_1\gamma_2\gamma_3$ squares to $1$ and commutes with all the generators of $Cliff_3$. Hence the matrix representation of the element $\gamma_1\gamma_2\gamma_3$ can be used as a candidate to fill up the first column. Since we have now filled up two matrices (including the identity matrix) in the first column, it should be possible to get a $2$-real symbol decodable code using matrix representation of $Cliff_3$. From Section \ref{sec2}, we know that the remaining weight matrices should be obtained as a product of matrices in the first row and those in the first column. We have,

{\footnotesize
\begin{equation*}
(\gamma_1)(\gamma_1\gamma_2\gamma_3)=-\gamma_2\gamma_3;~
(\gamma_2)(\gamma_1\gamma_2\gamma_3)=\gamma_1\gamma_3;~
(\gamma_3)(\gamma_1\gamma_2\gamma_3)=-\gamma_1\gamma_2.
\end{equation*} 
}
It so turns out that the elements $\left\{1, \gamma_1, \gamma_2, \gamma_3, -\gamma_1\gamma_2, -\gamma_2\gamma_3, \gamma_1\gamma_3, \gamma_1\gamma_2\gamma_3\right\}$ also form a basis for $Cliff_3$. Thus a general element of $Cliff_3$ can be expressed as
\begin{equation*}
\begin{array}{rcl}
x&=&a_1+\gamma_1a_2+\gamma_2a_3+\gamma_3a_4\\
&&+(-\gamma_1\gamma_2)a_5+(-\gamma_2\gamma_3)a_6+(\gamma_1\gamma_3)a_7+\gamma_1\gamma_2\gamma_3a_8
\end{array}
\end{equation*}
for some $a_i\in\mathbb{R},i=1,\dots,8$. By thinking of the element $\gamma_1$ as the complex number $i=\sqrt{-1}$, we can view $Cliff_3$ as a vector space over $\mathbb{C}$. To be precise,
\begin{equation*}
\begin{array}{rcl}
x&=&(a_1+\gamma_1a_2)+\gamma_2(a_3+\gamma_1a_5)\\
&&+\gamma_3(a_4-\gamma_1a_7)+\gamma_2\gamma_3(-a_6+\gamma_1a_8)\\
&=&z_1+\gamma_2z_2+\gamma_3z_3+\gamma_2\gamma_3z_4
\end{array}
\end{equation*}
where, $z_i\in\mathbb{C},i=1,\dots,4$ and are given by

{\small
\begin{equation*}
\begin{array}{rcl}
z_1=(a_1+\gamma_1a_2); & & z_2=(a_3+\gamma_1a_5);~ \\
z_3=(a_4-\gamma_1a_7); & & z_4=(-a_6+\gamma_1a_8).
\end{array}
\end{equation*}
}
Now using left regular representation as in the case of codes from $\mathbb{A}_2^L$, we obtain the following design
\begin{equation*}
L_x=\left[\begin{array}{rrrr}
z_1 & -z_2^* & -z_3^* & -z_4\\  
z_2 & z_1^* & -z_4^* & z_3\\
z_3 & z_4^* & z_1^* & -z_2\\
z_4 & -z_3^* & z_2^* & z_1
\end{array}\right].
\end{equation*}

In general, for $R=2^\lambda$ relays we take the left regular representation of $\mathbb{A}_3^{2^{\lambda-2}}$.
\begin{eg}
Suppose we want a design for $R=8=2^3$ relays. Hence we have $\lambda=3$. Using the left regular representations of the algebra $\mathbb{A}_3^2,$  we get the following linear design

{\small
\begin{equation*}
L_x=\left[\begin{array}{rrrrrrrr}
z_1 & -z_2^* & -z_3^* & -z_4 & z_5 & -z_6^* & -z_7^* & -z_8\\  
z_2 & z_1^* & -z_4^* & z_3 & z_6 & z_5^* & -z_8^* & z_7\\
z_3 & z_4^* & z_1^* & -z_2 & z_7 & z_8^* & z_5^* & -z_6\\
z_4 & -z_3^* & z_2^* & z_1 & z_8 & -z_7^* & z_6^* & z_5\\
z_5 & -z_6^* & -z_7^* & -z_8 & z_1 & -z_2^* & -z_3^* & -z_4\\  
z_6 & z_5^* & -z_8^* & z_7 & z_2 & z_1^* & -z_4^* & z_3\\
z_7 & z_8^* & z_5^* & -z_6 & z_3 & z_4^* & z_1^* & -z_2\\
z_8 & -z_7^* & z_6^* & z_5 & z_4 & -z_3^* & z_2^* & z_1
\end{array}\right].
\end{equation*}
}

The corresponding $4$ groups of real variables are $\left\{z_{1I},z_{4Q},z_{5I},z_{8Q}\right\}$, $\left\{z_{1Q},z_{4I},z_{5Q},z_{8I}\right\}$, $\left\{z_{2I},z_{3Q},z_{6I},z_{7Q}\right\}$ and $\left\{z_{3I},z_{2Q},z_{7I},z_{6Q}\right\}$.
\end{eg}

\section{Conjugate-linearity, unitary relay matrices and full-diversity}
\label{sec6}
Note that both the classes of codes from $\mathbb{A}_2^L$ and $\mathbb{A}_3^L$ have the property that any column of the design has only the variables or their conjugates. This is by virtue of the properties of left regular representation. While taking the left regular matrix representation, recall that we viewed the algebra as a vector space over $\mathbb{C}$ by thinking of the element $\gamma_1$ as the analogue of the complex number $i=\sqrt{-1}$. Any column of the design was then obtained as the image of a few elements of the natural basis of the algebra under the map $L_x$. All the elements of the natural basis of $\mathbb{A}_n^L$ have the property that they either commute with $\gamma_1$ or anticommute with $\gamma_1$. When we found the image of a basis element say $y$, recall that we moved $y$ past a complex number $z_i$. If $y$ commutes with $\gamma_1$, then it leaves the complex number intact. If $y$ anticommutes with $\gamma_1$, then it inflicts conjugation while moving past the complex number. This fact can be clearly observed in \eqref{eqn_matrix}. 

Moreover, it can be easily observed that all the relay matrices of the resulting designs are unitary. This is because the number of complex variables in the design is equal to the size of the matrix and by virtue of the left regular representation any complex variable appears only once in any column. Further, the positions in which they appear in different columns is different.       

Full diversity can be obtained for all the constructed codes by choosing an appropriate rotated $\mathbb{Z}^{\frac{R}{2}}$ lattice constellation. This has been proved in \cite{KaR3} more generally for all the codes constructed using the sufficient conditions for low ML decoding complexity discussed in Section \ref{sec2}. 

\section*{Acknowledgment}
This work was supported through grants to B.S.~Rajan; partly by the
IISc-DRDO program on Advanced Research in Mathematical Engineering, and partly
by the Council of Scientific \& Industrial Research (CSIR, India) Research
Grant (22(0365)/04/EMR-II).


\end{document}